\documentclass[conference]{IEEEtran}
\IEEEoverridecommandlockouts

\usepackage{import}
\usepackage{preamble}

\def\BibTeX{{\rm B\kern-.05em{\sc i\kern-.025em b}\kern-.08em
		T\kern-.1667em\lower.7ex\hbox{E}\kern-.125emX}}

\begin{document}

\title{
	{\LARGE \textsc{\\Measure of Dependence for Financial Time-Series}}\\
	\vspace{-5pt}
}

\author{

	\IEEEauthorblockN{Martin Winistörfer}
	\IEEEauthorblockA{
	\textit{Equality Fund}\\
	martin.winistoerfer@equalityfund.ch}
	\vspace{-15pt}
\and

	\IEEEauthorblockN{Ivan Zhdankin}
	\IEEEauthorblockA{
	\textit{Machine Learning Institute}\\
	ivan.zhdankin@thalesians.com}
	\vspace{-15pt}
}

\maketitle
\thispagestyle{plain}
\pagestyle{plain}

\begin{abstract}
Assessing the predictive power of both data and models holds paramount significance in time-series machine learning applications. Yet, preparing time series data accurately and employing an appropriate measure for predictive power seems to be a non-trivial task. This work involves reviewing and establishing the groundwork for a comprehensive analysis of shaping time-series data and evaluating various measures of dependence. Lastly, we present a method, framework, and a concrete example for selecting and evaluating a suitable measure of dependence.
\end{abstract}

\begin{IEEEkeywords}
Time-series, finance, predictive power, predictability, data shaping, feature importance, data preprocessing
\end{IEEEkeywords}

\section{Introduction}

To develop a successful machine learning application, it is crucial to identify the data with the highest predictive power. The machine learning model can then extract information from the data and make accurate predictions. Identifying such data is part of the data preparation step for a typical machine learning pipeline.

In financial applications, a model could predict asset movement for equities, balance portfolio allocations, predict bond spreads, manage risk, or execute limit book orders. In all cases, financial time series data is used as input data, typically market data such as price data. Depending on the problem statement, alternative datasets can be of high value, including data from social media, company news headlines (sentiment), economic activities, or fundamental data.

Additionally, deep neural networks are deployed for financial applications, with complex architectures using many layers of convolutional, recurrent, fully connected layers, or ensembles of models. These models have numerous hyperparameters, and their training is computationally expensive and time-consuming.

Given the multitude of data sources and the complexity of models, developing an optimal system is a challenging task. It is desirable to separate this high-dimensional search space for an optimal solution. Specifically, it would be highly beneficial to separate the identification of data with high predictability from the hyperparameter search and training process. By doing so, it would be possible to preselect potential data with high predictive power and later extract the detailed granularity of predictive power using the production model and training process.

For this purpose, we need measures of dependence between the input data and the dependent data. In a supervised machine learning application, the dependent data is the target respectively ground truth label.

The optimal measure of dependence is a metric that gives an indication of the degree of information contained in the input data to predict the target. A value of 0 could indicate that there is no relevant information in the input data, and the dependence between input and target is purely random. A value of 1 could indicate high dependence of input data on the target. With such data, a machine learning production model could extract the relevant information and make accurate predictions.

The Pearson correlation is a concrete example of such a measure of dependence. It compares the input data $x$ with output data $y$\footnote{Symbols with lowercase letters denote vector data, and symbols with uppercase letters denote matrices or multi-dimensional tensors.} and outputs a correlation coefficient. Pearson correlation is based on a normalized covariance calculation and is a well-known statistical method to find relationships between two variables.

When looking for an optimal measure of dependence, it is important to differentiate between purely model-free measures of dependence and model-based measures. Model-based measures utilize a model explicitly to calculate a loss, such as a decision tree, a naive prediction model, or even a neural network. In any case, such measures always include the properties of a model in the output. For instance, the dependency between input data and the target could be purely random. However, depending on the data, the model might overfit, and the loss is almost zero. In such a case, the measure would indicate a high value of dependence, even though there is, in fact, no dependence. The Pearson correlation would be a model-free measure using a model implicitly by the calculation of the coefficient, which is, in essence, a mapping of $x$ to $y$ data.

Apart from finding the optimal measure of dependence, the challenge is to match such a measure with time series data. In most financial applications, the Markov process does not hold true. Consequently, a machine learning model must be fed with historic data. At the same time, financial data can contain many features. Thus, we have to deal with two-dimensional data where the main dimension is time, and the other dimension is the features.

The time dimension is strictly sequential, with the data order fixed in time. Additionally, depending on the problem statement, the data can contain additional dimensions, such as different assets.

\begin{figure}[h]
	\centering
	\resizebox{0.4\textwidth}{!}{\noindent\begin{tikzpicture}

\draw[draw=black] (0,0) rectangle ++(2.5,0.8);
\node at (1.25,0.55) {X};
\node at (1.25,0.2) {\tiny (samples, time steps, features)};

\draw[draw=black] (3,0) rectangle ++(2,0.8);
\node at (4,0.55) {Y};
\node at (4,0.2) {\tiny (samples, time horizon)};

\end{tikzpicture}}
	\caption{Time series data transformed into two tensors $X$ and $Y$. In this example $X$ is of rank 3. The rank could be higher. For instance, instead of having features for one stock only, each stock could have its own feature. In such a scenario $X$ would be of rank 4.}
	\label{fig_format}
\end{figure}
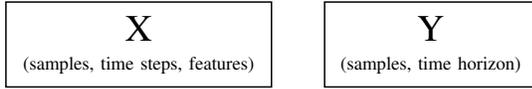

In supervised machine learning, time series data is typically shaped into a tabulated format using a sliding window. Each row contains all information about one time step (sample). The columns contain the features for each time lag and targets for each time horizon. Nevertheless, it is more convenient to think in tensors instead. Figure \ref{fig_format} depicts two tensors $X$ and $Y$ for a multivariate and multi-step problem statement. $Y$ includes a time horizon.

In a multi-step problem statement, the goal is not only to make a single prediction into the future but to predict multiple future values at once. This constitutes a typical time series forecasting task. For this purpose, the machine learning model is fed with multiple future target values. Such a model has multiple outputs to predict the time horizon and could, for instance, consist of LSTM network architecture.

In financial trading applications, technical analysis is a common tool for trading assets such as stocks. A technical analyst would use price trends, patterns, or indicators such as momentum or volume indicators to predict future prices. Essentially, a technical analyst or even a long-term investor tries to identify regimes in the data that might persist over time well into the future. Like an analyst or investor, the general goal is to detect regimes.

It's important to understand that a regime cannot be detected with a single-step prediction, i.e., one future time step. While it is possible to identify a pattern or regime in historical data, it is highly unlikely that there is a dependence between such a regime and a single data point in the future.

Therefore, in this paper, we always consider a time horizon with the ultimate goal of detecting a regime that persists over time from historical data through future data.

\begin{figure}[h]
	\centering
	\resizebox{0.5\textwidth}{!}{\noindent\begin{tikzpicture}

\tikzstyle{every node}=[font=\large]

\node at (4,4.9) {Historic};
\draw [decorate, decoration = {brace, amplitude=3.5mm}] (2,4.2) -- (6,4.2);
\node at (8,4.9) {Time Horizon};
\draw [decorate, decoration = {brace, amplitude=3.5mm}] (6,4.2) -- (10,4.2);

\draw[nice table] (0,0) grid (12,4); 

\node at (-0.4,0.5) {$f_4$}; \node at (-0.4,1.5) {$f_3$}; \node at (-0.4,2.5) {$f_2$};\node at (-0.4,3.5) {$f_1$};

\filldraw[blue cell] (2,0) rectangle (6,4);
\filldraw[red cell]  (6,0) rectangle (10,4);

\draw[time arrow] 	(-1,-0.8) -- (-1,5) 	node[anchor=south] {Probability};
\draw[time arrow] 	(-1,-0.8) -- (12.5,-0.8) 	node[anchor=north] {Time};
\draw[dotted]   	(6,0)    -- (6,-1) 		node[anchor=north] {$t_0$};
\draw[dotted]   	(7,0)    -- (7,-1) 		node[anchor=north] {$t_1$};
\draw[dotted]   	(9,0)    -- (9,-1) 		node[anchor=north] {$t_3$};
\draw[dotted]   	(4,0)    -- (4,-1) 		node[anchor=north] {$t_{-2}$};
\draw[thick]  		(6,0)    -- (6,4);

\draw [ultra thick, blue] plot [smooth, tension=0.8] coordinates {(3.4,3.5) (3.6,2.5) (3.8,3.5) (4,2.5) (4.2,3.5) (4.4,2.5) (4.6,3.5)};
\draw [ultra thick, red] plot [smooth, tension=0.8] coordinates {(8.4,3.5) (8.6,2.5) (8.8,3.5) (9,2.5) (9.2,3.5) (9.4,2.5) (9.6,3.5)};
\draw [ultra thick, dashed] plot [smooth, tension=0.8] coordinates {(0,-0.8) (3.5,2) (6,0.5) (8.6,2) (12,-0.8)};

\end{tikzpicture}}
	\caption{Exemplary time series data with relevant information embedded in the historic and future data points. The dashed line indicates the probability of occurrence of relevant features along the time axis. }
	\label{fig_regime}
\end{figure}
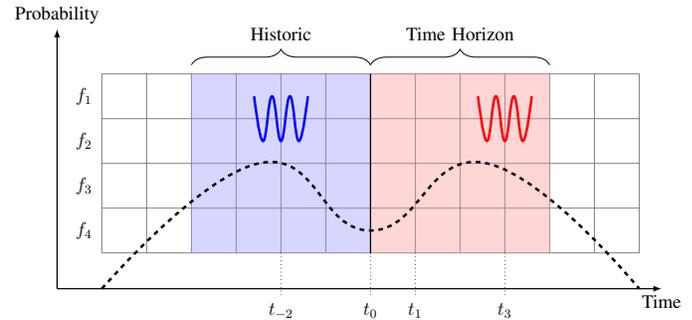

Figure \ref{fig_regime} illustrates an time series with four features. The input data is in blue, the future data is marked red. The time horizon consists of the following four time data points up to $t_{4}$.

At $t_{3}$, there is a piece of relevant future information as part of a regime in the data. If the model could recognize this piece of information, it would be able to predict the future asset price and make profitable trades. This piece of relevant future information has a strong dependency on another piece of information in the past at $t_{-2}$.  A measure of dependence would need to consider both pieces of information to measure the dependence accurately. If the measure of dependence only knew the piece of information at $t_{-2}$ and a hypothetical piece of information at $t_1$, it would detect no dependence. The detection of a regime must be symmetrical between historical and future data.

Moreover, the dashed line shows the probability of relevant information in the data over time. Very old data or data far in the future most likely have a low probability of containing relevant information. The regimes have faded or have not yet revealed themselves. Whereas very short time data points might not be able to pick up regimes and are of a random, noisy nature. Consequently, the probability that this data has relevant information drops. The maximum must be in between, as indicated by the maximum bumps in probability in Figure \ref{fig_regime}.

It is the goal of this paper to present an overview of the measures of dependence that are applicable for time series data, particularly for financial data, but could be deployed for other time series data such as sensory data.

In the coming sections, we describe the different ways in which to best prepare the time series data for the different measures of dependence. Afterwards, we provide an overview of the applicable measures and explain each measure in more detail. At the end of this paper, we conduct an experiment on real stock market data and compare the results of the different measures. It is not the goal of this paper to identify the best measure of dependence, but to give the theoretical basis to evaluate and select the best measure for a particular application.

\section{Time Series Shaping}

To use financial time series data in a machine learning model, it is necessary to transform the data. For a supervised machine learning problem, data is very often in a tabular form, making it easy to process. Many machine learning libraries and tools require data to be in a matrix format. However, this is not a strict requirement, as neural network training with supervised learning can accept tensors as inputs, as outlined in the introductory section. In fact, with reinforcement learning setups, data is very often not in tabular form but in multidimensional tensors.

\begin{table}[h]
	\caption{Time series transformation in to two data formats.}
	\begin{center}
		\renewcommand{\arraystretch}{1.5}
		\begin{tabular}{ |p{1.5cm}|p{6cm}| }
			\hline Data Format & Description\\
			\hline X:Y & Measures between a vector of input and a vector of output data (uni-variate-multi-step). \\
			\hline S:Y & Measures between as set of vector of inputs $S = \set{X_1, ... X_N}$ and a vector of output data (multivariate-multi-step). \\
			\hline
		\end{tabular}
		\label{tab_transformation}
	\end{center}
\end{table}

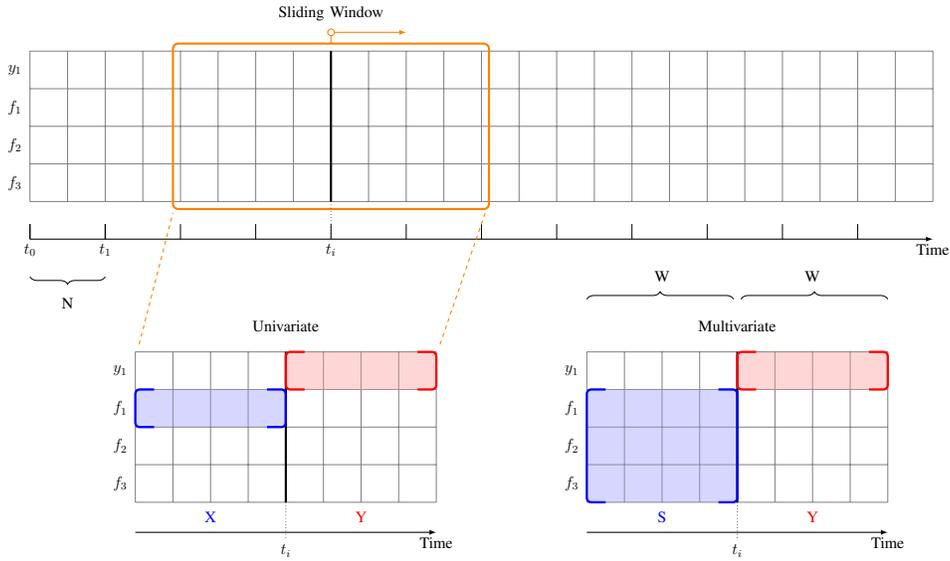
\begin{figure*}[h]
	\centering
	\resizebox{0.7\textwidth}{!}{\noindent\begin{tikzpicture}

\tikzstyle{every node}=[font=\large]

\node at (8,5) {Sliding Window};
\draw[nice table] (0,0) grid (24,4); 

\node at (-0.4,0.5) {$f_3$}; \node at (-0.4,1.5) {$f_2$}; \node at (-0.4,2.5) {$f_1$};\node at (-0.4,3.5) {$y_1$};

\draw[rounded corners, ultra thick, orange] (3.8, -0.2) rectangle (12.2, 4.2) {};
\draw[orange, thick]	(8,4.2)   -- (8,4.4);
\draw[blue arrow, orange] 	(7.9,4.5) -- (10,4.5);

\draw[time arrow] 	(0,-1) -- (24,-1) 	node[anchor=north] {Time};
\foreach \x in {0,2,...,22}{ \draw (\x,-0.6) -- (\x,-1); }
\draw[thick]		(0,-0.6) -- (0,-1) 		node[anchor=north] {$t_0$};
\draw[thick]   		(2,-0.6) -- (2,-1) 		node[anchor=north] {$t_1$};
\draw[dotted]   	(8,0)    -- (8,-1) 		node[anchor=north] {$t_{i}$};
\draw[ultra thick]  (8,0)    -- (8,4);

\node at (1,-2.7) {N};
\draw [decorate, decoration = {brace, amplitude=2mm, mirror}] (0,-2) -- (2,-2);

\tikzset{shift={(12,0)}}

\tikzset{shift={(-9.2,-8)}}

\node at (4,4.7) {Univariate};
\draw[nice table] (0,0) grid (8,4); 
\node[blue] at (2,-0.4) {X}; \node[red] at (6,-0.4) {Y};

\node at (-0.4,0.5) {$f_3$}; \node at (-0.4,1.5) {$f_2$}; \node at (-0.4,2.5) {$f_1$};\node at (-0.4,3.5) {$y_1$};

\filldraw[blue cell] (0,3) rectangle (4,2);
\filldraw[red cell]  (4,3) rectangle (8,4);

\draw[thick,->, -latex] (0,-0.8) -- (8,-0.8) 	node[anchor=north] {Time};
\draw[dotted]   		(4,0)    -- (4,-1) 		node[anchor=north] {$t_i$};
\draw[ultra thick]  	(4,0)    -- (4,4);

\draw [square brace] 	(0.5,2) to [square left brace ] (0.5,3);

\draw [square brace] 	(3.5,2) to [square right brace] (3.5,3);

\draw [square brace, red] (4.5,3) to [square left brace] (4.5,4);

\draw [square brace, red] (7.5,3) to [square right brace] (7.5,4);

\draw[dashed, orange, thick]  	(0.1,4.2)   -- (1,7.7);
\draw[dashed, orange, thick]  	(8.1,4.2)   -- (9.3,7.7);

\tikzset{shift={(12,0)}}

\node at (4,4.7) {Multivariate};
\draw[nice table] (0,0) grid (8,4); 
\node[blue] at (2,-0.4) {S}; \node[red] at (6,-0.4) {Y};

\node at (-0.4,0.5) {$f_3$}; \node at (-0.4,1.5) {$f_2$}; \node at (-0.4,2.5) {$f_1$};\node at (-0.4,3.5) {$y_1$};

\filldraw[blue cell] 	(0,0) rectangle (4,3);
\filldraw[red cell]  	(4,3) rectangle (8,4);

\draw[thick,->, -latex] (0,-0.8) -- (8,-0.8) 	node[anchor=north] {Time};
\draw[dotted]   		(4,0)    -- (4,-1) 		node[anchor=north] {$t_i$};
\draw[ultra thick]  	(4,0)    -- (4,4);

\draw [square brace] 	(0.5,0) to [square left brace ] (0.5,3);

\draw [square brace] 	(3.5,0) to [square right brace] (3.5,3);

\draw [square brace, red] (4.5,3) to [square left brace ] (4.5,4);

\draw [square brace, red] (7.5,3) to [square right brace] (7.5,4);

\node at (2,6) {W};
\draw [decorate, decoration = {brace, amplitude=2mm}] (0,5.4) -- (3.9,5.4);
\node at (6,6) {W};
\draw [decorate, decoration = {brace, amplitude=2mm}] (4.1,5.4) -- (8,5.4);

\end{tikzpicture}}
	\caption{Overview of the $X:Y$ and $S:Y$ time series data transformation using a sliding window with a window width of $W=4$ and a step size of $N=2$. Data is processed from left to right. The time series sequence potentially represents the full data set or part of a data sequence like when using cross validation.}
	\label{fig_data}
\end{figure*}

In all cases, a type of sliding window is used to transform the continuous stream of time series data into processable units. This transformation is accomplished either by preprocessing the data or by generating the data units ad-hoc during the training process. In this paper, we focus on a generalized form of data that can be best used as inputs for a particular measure of dependence. Table \ref{tab_transformation} lists the two data formats. As shown in the table, the target variable is always a single vector.

Both transformations are visualized in Figure \ref{fig_data}. The example in Figure \ref{fig_data} has three features and one target. As depicted, the target is a different time series than the features. This definition differs from typical time series forecasting setups where the goal of the forecast is to predict the same time series into the future. Also, one could imagine having multiple time series targets instead of just one.

\section{Overview of Measures}

\subsection{Statistical Methods}

Statistical methods are the most common way to compare data and find relationships or associations between two variables. During data exploration and understanding phase, covariance matrix or correlation coefficients are used by data scientists to understand the data, spot data specific patterns and make selective choices about which data to use. Most of these measures work on pairwise data only and a major drawback of these measures is that they often can only detect linear correlations. However, in financial data it is assumed that relevant data with high predictive power is only weakly embedded in the input data and of non-linear nature.

\subsection{Information Theory Measures}

Since its introduction by Shannon to quantify communication, information theory \cite{commenges} has proven to be a valuable tool in many disciplines. Information theory's broad applicability is partly due to its reliance solely on the probability distribution associated with one or more variables. In general, information theory utilizes the probability distributions linked to the values of variables to determine whether or not these values are related and, depending on the context, the nature of their relationship. The most fundamental information theory measures we examine here are entropy and mutual information. Similar to statistical methods, information theory employs an implicit model to map $x$ and $y$ data. However, because we do not explicitly specify a model, these measures are considered model-free. In principle, data alone cannot predict other data without some form of a model.

\subsection{Model-Based Methods}

Model-based methods use a model explicitly, such as a decision tree. Even a machine learning production model using a neural network could be considered as a measure of dependence that uses a model explicitly. In any case, it is important to understand the problem statement at hand. The main questions to be answered by a true measure of dependence are:

\begin{enumerate}
	\item How much information is in the input data that can explain the target?
	\item Is there a regime and a relationship between past and future values?
\end{enumerate}

Most model-based methods ask a different question: "How well can a model extract information and make an accurate prediction?" Such a question focuses on the construction of the model, not the data. Finding relevant information or regimes in the data is an implicit requirement of the model. It assumes that the model can do it by default, which depends on the construction of the model. It is preferred to remove this additional uncertainty of model construction from a measure of dependence.

Model-based methods are compelling because they are closer to the final implementation of the machine learning algorithm. If the model of the measure of dependence were the same as the production model, it would be best not to preselect features with high predictive power but evaluate in a single step.\label{chap_model_based_methods} If a production model is complex, it would be advantageous to use a simpler model for the measure of dependence. In that case, the result could be that such a measure would not detect the relevant information in the data or the wrong data, potentially invalidating the training process.

\subsection{Synergy}

Synergy is an important concept when looking at measure of dependence \cite{timme}. To begin to understand synergy, we can use a simple system. Suppose two variables (call them $X_1$ and $X_2$) provide some information about a third variable (call it $Y$). In other words, if you know the state of $X_1$  and $X_2$ , then you know something about the state Y . Loosely said, the portion of  information that is not provided by knowing both $X_1$ alone and $X_2$ alone is said to be provided synergistically by $X_1$  and $X_2$ . The synergy is the bonus information received by knowing $X_1$  and $X_2$ together, instead of separately.

Therefore the ideal measure of dependence can concurrently look at the input data $X$ and compare to the target $Y$. As we will see later, most of the presented measure, can actually only compare between one input variable and one output variable at the same time and ignore the potential of synergy in the input data.

This is particularly critical for financial time series in which a combination of input data could yield a new signal for machine learning trading algorithm. Consequently, synergy is an important concept for financial time series predictions. 

\subsection{Type of Interactions}

When looking at measure of dependence one must pay specific attention to two type of interactions: 1) those that exist within a group of variables and 2) those that exist between a
group of variables and another target variable. This is called “Group to Target”. In such applications we always have labeled data. For unsupervised machine learning or forecasting tasks, the first type of measure could be used. For machine learning prediction application only the last type of interaction is of importance.

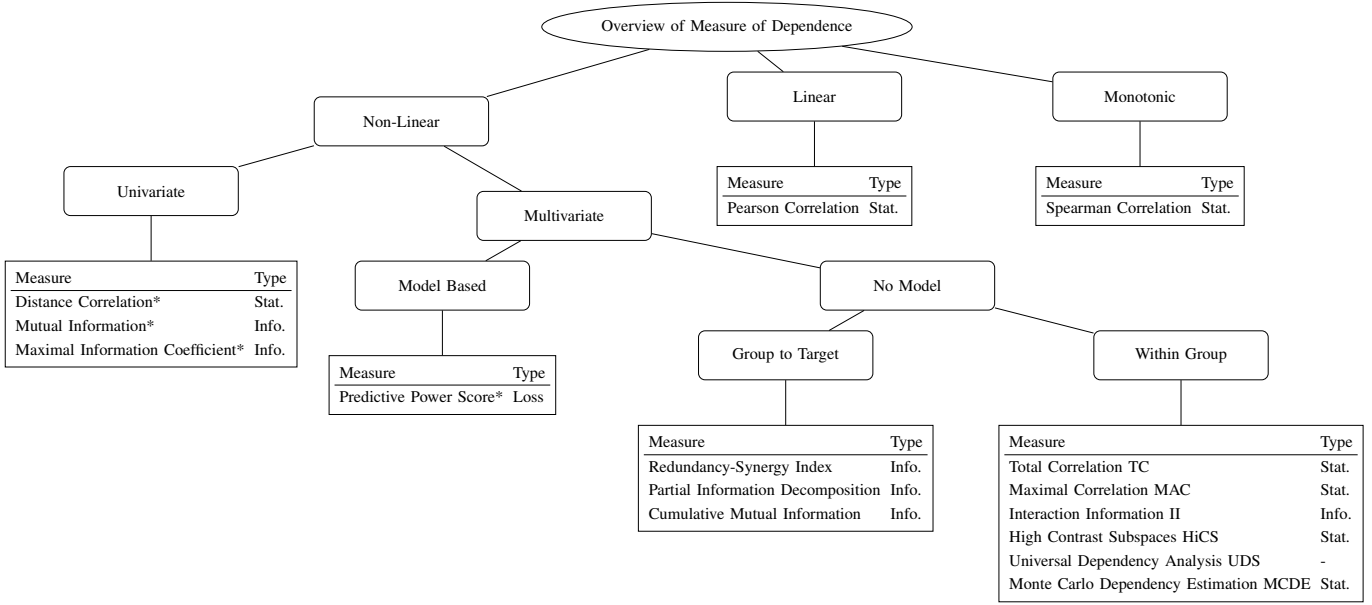
\begin{figure*}[tb]
\centering
\resizebox{1.0\textwidth}{!}{\begin{tikzpicture}

\tikzset{
    boxnode/.style={rectangle,
        rounded corners,
        draw=black, 
        text width=10em,
        minimum height=3em,
        text centered
    },
    endnode/.style={rectangle,
		draw=black, 
		minimum height=3em
	},
	titlenode/.style={ellipse,
		draw=black, 
		minimum height=3em
	},
}

\tikzstyle{level 1}=[sibling distance=70mm]
\tikzstyle{level 2}=[sibling distance=70mm]
\tikzstyle{level 3}=[sibling distance=90mm]
\tikzstyle{level 4}=[sibling distance=110mm]

\tikzstyle{every node}=[anchor=north]

\node [titlenode] (param) {Overview of Measure of Dependence}
    child { node [boxnode] (non-linear) {Non-Linear}
		child { node [boxnode, left=0mm] (univariate) {Univariate}
		   	child { node [endnode] 
		   		{	\begin{tabular}{ll}
		   			Measure & Type \\
		   			\hline
		   			Distance Correlation* & Stat.\\
		   			Mutual Information* & Info.\\
		   			Maximal Information Coefficient* & Info.\\
		   			\end{tabular}
		   		}	
		   	}
		}
    	child { node [boxnode] (multivariate)  {Multivariate}
			child { node [boxnode, right=0mm] (model) {Model Based}
				child { node [endnode]
					{	\begin{tabular}{ll}
						Measure & Type\\
						\hline
						Predictive Power Score* & Loss\\
						\end{tabular}
					}
				}
			}
			child { node [boxnode, right=10mm] (nomodel) {No Model}
				child { node [boxnode, right=10mm] (grouptotarget) {Group to Target}
					child { node [endnode]
						{	\begin{tabular}{ll}
							Measure & Type\\
							\hline
							Redundancy-Synergy Index & Info. \\
							Partial Information Decomposition & Info.\\
							Cumulative Mutual Information & Info.\\
							\end{tabular}
						}	
					}
				}
				child { node [boxnode, right=-15mm] (within) {Within Group}
					child { node [endnode]
						{	\begin{tabular}{ll}
							Measure & Type\\
							\hline
							Total Correlation TC & Stat.\\
							Maximal Correlation MAC & Stat.\\
							Interaction Information II& Info.\\
							High Contrast Subspaces HiCS & Stat.\\
							Universal Dependency Analysis UDS & -\\
							Monte Carlo Dependency Estimation MCDE & Stat.\\
							\end{tabular}
						}	
					}
				}
			}
		}
    }
	child { node [boxnode, right=0mm] (Linear){Linear}
    	child { node [endnode] 
    		{	\begin{tabular}{ll}
    			Measure & Type \\
    			\hline
    			Pearson Correlation & Stat.\\
    			\end{tabular}
    		}	
    	}
    }
	child { node [boxnode, right=0mm] (monotonic){Monotonic}
		child { node [endnode] 
			{	\begin{tabular}{ll}
				Measure & Type \\
				\hline
				Spearman Correlation & Stat.\\
				\end{tabular}
			}	
		}
	}
;

\end{tikzpicture}}
\caption{An overview of measures of dependence grouped by relevant categories. The metrics itself are grouped into three types: statistical metrics (stat), information theory based metrics (info) and model based metrics (loss).}
\label{fig_metrics}
\end{figure*}

Figure \ref{fig_metrics} presents an overview of the measures of dependence that have been considered in this work.

\section{Measure of Dependence}

This section gives a short introduction to each measure of dependence. 

\subsection{Pearson Correlation PC}

In statistics, the Pearson correlation coefficient is a measure of linear correlation between two sets of data and is defined as:

\begin{equation}
\rho = \frac{cov_{xy}}{\sigma_x \sigma_y} \label{eq_pearson}
\end{equation}

such that

\begin{equation}
cov_{xy} = \sum^N_i\frac{(x_i - \hat{x})(y_i - \hat{y})}{N-1} \label{eq_pearson_2}
\end{equation}

\subsection{Spearman Correlation (SC)}

The Spearman rank-order correlation coefficient, is a nonparametric measure of the strength and direction of association between two variables. Unlike Pearson's correlation, which assesses linear relationships, Spearman's correlation evaluates monotonic relationships. Monotonic relationships are those where the variables tend to move in the same relative direction, but not necessarily at a constant rate. Spearman's correlation is particularly useful when the relationship between variables is more accurately described by their ranks rather than their exact values.

\begin{equation}
r_s = \rho_{R(X),R(Y)}  = \frac{cov(R(X), R(Y))}{\rho_{R(X)}\rho_{R(Y)}} \label{eq_spearman}
\end{equation}

where:

\begin{equation_where}
	\rho & denotes the usual Pearson correlation coefficient, but applied to the rank variables, \\
	cov(R(X), R(Y)) & is the covariance of the rank variables,\\ 
	R(X), R(Y) & are the standard deviations of the rank variables.
\end{equation_where}

The Spearman correlation is less sensitive than the Pearson correlation to strong outliers located in the tails of both samples. This is due to Spearman's coefficient limiting the impact of outliers to the value of their rank. When the data are approximately elliptically distributed and devoid of prominent outliers, the Spearman correlation and Pearson correlation yield similar values.

\subsection{Distance Correlation DC}

Differing from classical measures such as Pearson or Spearman correlation, distance correlation is a correlation measure that captures both linear and non-linear associations between two random variables, $X$ and $Y$ \cite{szekely}. Distance correlation equals zero if, and only if, two vectors are independent. This characteristic makes distance correlation a statistical measure of independence between random variables \cite{wiki_distance}.

Distance correlation, also referred to as Brownian covariance or correlation, is defined by the following coefficient:

\begin{equation}
dCor(X,Y) = \frac{dCov(X,Y) }{\sqrt{dVar(X)\; dVar(Y) }} \label{eq_dc}
\end{equation}

The distance correlation is derived from a number of other quantities that are used in this definition, specifically: distance variance, distance standard deviation, and distance covariance \footnote{In Python, for easy computation of sample distance correlation the \emph{dcor()} library can be used \cite{dcor}.}. 

\subsection{Maximum Information Coefficient MIC}

The maximal information coefficient is a measure that is based on information theory particularly mutual information\cite{robidoux}. The maximal information coefficient is an approach to try many different bin sizes and locations, and to compare the maximum mutual information received. It is defined as 

\begin{equation}
MIC \triangleq max_{x,y:xy<B} m(x,y) \label{eq_mic}
\end{equation}

such that

\begin{equation}
m(x,y) = \frac{max_{G\in G(x,y)}I(X(G), Y(G))}{log_2\min(x,y)} \label{eq_mic_m}
\end{equation}

where:

\begin{equation_where}
	B & is some sample-size dependent bound on the number of bins that can be used to reliably estimate the distribution, \\
	G(x,y) & is the set of two-dimensional grids of size $x × y$ and $X(G)$, $Y(G)$ represents a discretization of the variables onto this grid. \\
\end{equation_where}

The MIC lies in a range [0,1], where 0 represents no relationship between variables and 1 represents a noise-free relationship of any form, not just linear\footnote {In Python to calculate MIC score  \href{https://minepy.readthedocs.io/en/latest/python.html}{Minepy} library can be used.}. MIC will not give any indication of the type of the relationship, though. It is possible with the MIC to find interesting relationships between variables in a way that simpler measures, such as the correlation coefficient, cannot. With MIC the goal is equitability: similar scores will be seen in relationships with similar noise levels regardless of the type of relationship. 

Because of this, it may be particularly useful with high dimensional settings to find a smaller set of the strongest correlations. Where distance correlation might be better at detecting the presence of (possibly weak) dependencies, the MIC is more geared toward the assessment of strength and detecting patterns that we would pick up via visual inspection. 

Hence, it cannot handle a system of multivariate inputs and outputs and correlate it as an abstract structure. Other more recent MIC variants are BackMIC and ChiMIC \cite{cao}.

\subsection{Mutual Information MI}

The information theoretic quantities involving one and two variables are well-defined and their results are well understood. Regarding the probability distribution of one variable, call it $p(x)$, the canonical measure is the entropy $H(x)$ \cite{cover}. The entropy is given by:

\begin{equation}
H(X) = \sum^n_{i=1} P(x_i)\log_2P(x_i)  \label{eq:entropy}
\end{equation}

The entropy quantifies the amount of uncertainty that is present in the probability distribution. If the probability distribution is concentrated near one value, the entropy will be
low. If the probability distribution is uniform, the entropy will be maximized. When examining the relationship between two variables, the mutual information \ref{eq_mi} quantifies the amount of information provided about one of the variables by knowing the value of the other. The mutual information is given by:

\begin{equation}
\begin{split}
I(X;Y) 	&= H(X) - H(X|Y) \\
&= H(Y) - H(Y|X) \\
&= H(X) + H(Y) - H(X|Y)
\end{split}
\label{eq_mi}
\end{equation}

where the conditional entropy is given by:

\begin{equation}
\begin{split}
H(X|Y) 	&= \sum_{y \in Y} p(y)H(X|y) \\
&= \sum_{y \in Y} p(y) \sum_{x \in X} p(x|y)\log{\frac{1}{p(x|y)}} \\
\end{split}
\label{eq:mi_cond}
\end{equation}

where:

\begin{equation_where}
	X,Y & data variables, \\
	x,y & individual values of those variables.\\
\end{equation_where}

The mutual information can be used as a multivariate input measure of the interactions among more than two or more variables by grouping the variables into sets and treating each set as a single vector-valued variable. In this way, the mutual information can be used to measure the interactions between a group of variables and a target variable. For instance, the mutual information can be calculated between $Y$ and the set $S = \set{X_1, X_2}$. 

\subsection{Redundancy-Synergy Index RSI}

This is another multivariate information measure \cite{chechik} and  was created
as an extension of the interaction information \cite{mcgill}. It is given by:

\begin{equation}
RSI(S;Y)  \equiv I(S;Y) - \sum_{X_i \in S} I(X_i;Y) \label{eq_rsi}
\end{equation}

The redundancy-synergy index is designed to be maximal and positive when the variables in $S$ are purported to provide synergistic information about $Y$ . It should be negative when the variables in $S$ provide redundant information about $Y$. The redundancy-synergy index measures the interactions between a group of variables and another variable, except when $S$ contains two variables, in which case the redundancy-synergy index is equal to the interaction information. The redundancy-synergy index has been referred to as the SynSum \cite{globerson}, the WholeMinusSum synergy \cite{griffith and Koch}, and the negative of the redundancy-synergy index has also been referred to as the redundancy \cite{schneidman}.

Redundancy-Synergy Index RSI is one of the measures of dependence that naturally accepts multivariate inputs.

\subsection{Predictive Power Score PPS}

This measure is a model-based score that operates pairwise on $X$ and $Y$ data. The intuition behind this score is as follows:

A decision tree is fitted on the $X$ and $Y$ data, and the model's loss (Mean Absolute Error, MAE) is calculated. The loss is then normalized with the loss of a naive model. For a regression task, a naive approach would be to consistently use the median as the target. This way, approximately half of the values are expected to be above the naive prediction and half below. Subsequently, the model error is normalized with the error from the naive prediction.

\begin{equation}
PPS = 1 - \frac{MAE(y, \hat{y}_{model})}{MAE(y, \hat{y}_{median})} \label{eq_pps}
\end{equation}

where:

\begin{equation}
\hat{y}_{model} = f_{model}(x);\; \hat{y}_{median} = median(y) \label{eq_pps_where}
\end{equation}

In essence, the score measures the percentage of model improvement potential that the feature adds when the current naive base model is compared to a perfect deterministic model \cite{wetschoreck}.

The interesting thing, like any model based score, is that it can be easily extended to a multivariate problem setup having more than one $X$ data dimension \cite{vermunt}. This enables this score to detect combinations of feature variables. 

Like any model-based measure, the model can over-fit and the result of the score is ambiguous. This is particularly true, if data is not cross-validated \cite{molnar}. It is better to split data into a training and test data set, compute PPS on training set then measure dependence on test set. However, in most financial applications there is not enough data to execute such method, limiting its application.

\subsection{Information Gain IG}

Information gain is a synonym for Kullback–Leibler divergence and measures the relative entropy between two distributions \cite{wiki_ig}. It gives a measure of the amount of information gained about a variable observing another variable. The most prominent application of information gain is in decision trees. Information gain helps to determine the order of attributes in the nodes of a decision tree.  The information gain is the weighed average of all child nodes minus the parent node:

\begin{equation}
\begin{split}
IG(S, A) 	&= H(S)_{parent} - H(S,A)_{children} \\
&= H(S) - \sum_{t \in T} p(t)H(t)
\end{split}
\label{eq_ig}
\end{equation}

where:

\begin{equation_where}
IG &  information gain, the higher the information gain the better the split, the more entropy was removed from parent to children, \\
H(S) &  entropy of set S (parent),  \\   
T & the subsets created from splitting set $S$ attribute $A$ (child nodes) such that $S = \underset{t \in T}{\bigcup} t$, \\
p(t) & the proportion of the number of elements in $t$ to the number of elements in set $S$ (weighting of node entropies),\\
H(t) & entropy of subset $ $t (entropy of each child node).\\
\end{equation_where}

Information gain and Mutual Information are the same thing, although the context or usage of the measure often gives rise to the different names. For example:

\begin{itemize}
  \item Effect of transforms to a data set (\emph{decision trees}): Information Gain.
  \item Dependence between variables (\emph{feature selection}): Mutual Information.
\end{itemize}

Notice the similarity in the way that the information gain \ref{eq_ig} is calculated and the way that mutual information gain \ref{eq_mi} is calculated; when setting the parent data in a decision tree as $X$ and the child data as $Y$ they are equivalent. Technically, they calculate the same quantity if applied to the same data.

\section{Method} \label{chap_method}

There are many measures of dependence that could potentially be used to identify data with high predictive power. The question is, which one is the right one and how to verify it? Again, we must remember, that the identification of data with high predictive power is a preprocessing step before production model training.

\begin{figure}[h]
	\centering
	\resizebox{0.45\textwidth}{!}{\begin{tikzpicture}[node distance=2cm]

\node (start) [startstop, fill=gray!25] {Define the number of measure and feature space to be evaluate};
\node (step_1) [process,   below of=start, fill=gray!25] {Search for $M_{best,worst}$ and $F_{best,worst}$};
\node (step_2) [process,   below of=step_1] {Define production model and a subset of hyperparameters};
\node (step_3) [process,   below of=step_2] {Train the production model with $M_{best,worst}$ and $F_{best,worst}$};
\node (step_4) [process,   below of=step_3] {Correlate production model score $S_{highest,lowest}$ with $M_{best,worst}$ and $F_{best,worst}$};
\node (step_5) [decision,  below of=step_4, yshift=-1cm] {Positive Correlation?};
\node (stop)   [startstop, below of=step_5, yshift=-1cm] {Use $M_{best}$ and $F_{best}$ for full scale hyperparameter search};

\draw [arrow] (start) -- (step_1);
\draw [arrow] (step_1) -- (step_2);
\draw [arrow] (step_2) -- (step_3);
\draw [arrow] (step_3) -- (step_4);
\draw [arrow] (step_4) -- (step_5);
\draw [arrow] (step_5) -- node[anchor=south]  {no}  +(4,0) |- (start);
\draw [arrow] (step_5) -- node[anchor=east]  {yes} (stop);

\draw [decorate, decoration = {brace, mirror, amplitude=0.3cm}, thick] (-2.7,0.7) -- (-2.7,-2.7);

\node[align=right] at (-4.5,-1) {Algorithm to \\ find best measure \\ and features};

\end{tikzpicture}}
	\caption{Verification process to ensure, that measure of dependence and data have the required predictive power for a specific application. }
	\label{fig_verification}
\end{figure}
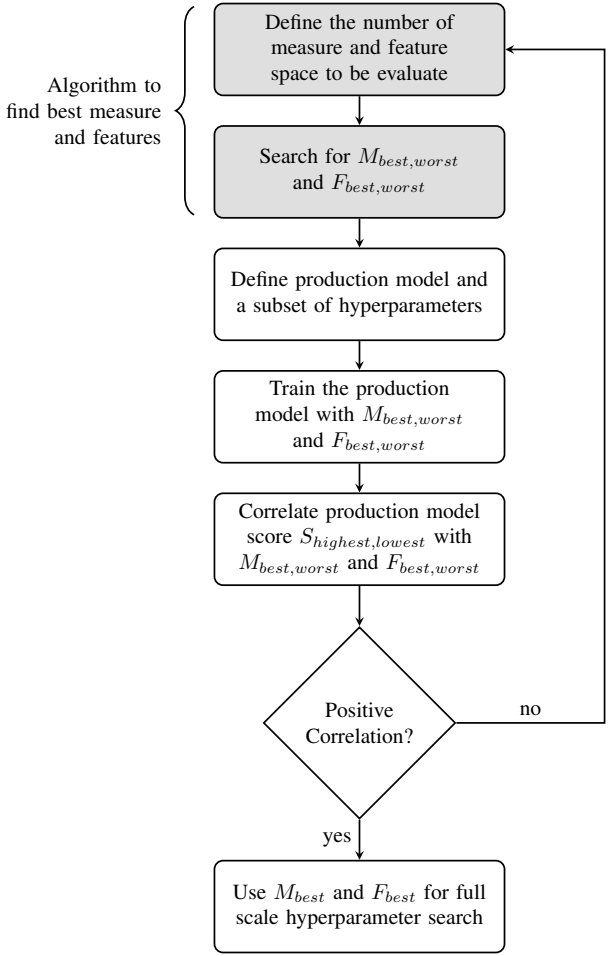

However, since the measure of dependence and the production model can deviate from each other, as outlined in Chapter \ref{chap_model_based_methods}, it becomes necessary to use the training results from the production model to judge the success of the preprocessing. It is assumed that this is an iterative process, and after a certain number of iterations, good measures and features can be distinguished from the less effective ones. The iterative verification process is outlined in Figure \ref{fig_verification}.

A positive correlation exists if the best measure $M_{best}$ and best feature $F_{best}$ values yield the highest production model score $S_{highest}$, and the worst measure $M_{worst}$ and worst feature $F_{worst}$ values yield the lowest production model score $S_{lowest}$. This correlation can be linear or of any monotonic nature. For cross-verification, an intermediate value can be used, or a range of values and scores can be compared.

This verification method is best applied with real data in an actual production environment. It is believed that this approach will yield optimal results. Alternatively, synthetic data can be generated. In such a scenario, signals, patterns, or even regimes can be embedded into the data. Purely random data can be used as a control dataset. In this setup, the data with high and low predictive power is already known in advance. The value of the measure of dependence would be expected to be higher with embeddings than with random data. However, especially in financial applications where patterns and regimes are not obvious, it would be very difficult to generate synthetic data. One could use generative models or, for specific applications like derivative investment instruments, a Black–Scholes model. Nevertheless, in this paper, the focus is on equity market predictions.

\section{Experiment}

We test three measure of dependence MI, DC and MIC and will conduct an experiment using stock market data. Specifically, the goal is to find features such as price returns, trading volume and technical indicators that have highest predict power\footnote {In Python a great library for technical indicators is \href{https://mrjbq7.github.io/ta-lib/}{TA-Lib.}}. Table \ref{tab_features} lists the targets and features that are used in this experiment. Price data is averaged using hull moving average HMA. This average minimized lag in the time series\footnote{See   \href{https://school.stock charts.com/doku.php?id=technical\_indicators:hull\_moving\_average}{HMA} for more details.}. Feature data are prefixed with f\_ and targets with t\_.

\begin{table}[h]
	\caption{List of features and targets used in the experiment.}
	\begin{center}
		\renewcommand{\arraystretch}{1.5}
		\begin{tabular}{ |p{2cm}|p{6cm}| }
			\hline Name & Description \\
			\hline f\_ret\_c\_1 & Daily closing price return \\
			\hline f\_vol\_pct & Daily stock trading volume change \\
			\hline f\_ret\_c\_hma\_5 & Avg. closing price return using 5-day HMA\\
			\hline f\_ret\_c\_hma\_20 & Avg. closing price return using 20-day HMA\\
			\hline f\_c\_macd\_signal & MACD signal\\
			\hline f\_c\_macd & MACD indicator\\
			\hline f\_obv & On balance volume oscillator\\
			\hline f\_atr & Average true range indicator\\
			\hline f\_rsi & Relative strength index\\
			\hline t\_ret\_c\_1 & Daily closing price return \\
			\hline t\_ret\_c\_hma\_5 & Avg. closing price return using 5-day HMA\\
			\hline t\_ret\_c\_hma\_20 & Avg. closing price return using 20-day HMA\\
			\hline
		\end{tabular}
		\label{tab_features}
	\end{center}
\end{table}

The targets are typically price returns. This can be daily returns t\_ret\_c\_1 or returns of any duration such as weekly or monthly. The target t\_ret\_c\_hma\_20 for example is a monthly average. The targets are computed before the search process at every time step.

The experiment is conducted with daily data from a single stock like YUM. The data ranges from 1999-11-01 to 2021-06-25.

It is important to note at this point, the duration of return and the sliding window size itself are key parameters in the search process. As outlined in figure \ref{fig_regime} regimes might exist over different time horizons. For each data the search process must find the optimal window. Also, the target is not a single value but a vector including data over a specific time horizon defined by the sliding window size.
 
Furthermore, the targets values are discretized into bins. In this example the bin size $B=10$ corresponds to 10 basis points (0.1\%). The bins size $B$ shall be chosen in resolution that makes sense for stock market financial investments.

\subsection{Algorithm}

The primary goal of the feature search algorithm is to find the right combination of parameters that maximizes the value of the measure of dependence. This is achieved through a grid search using Optuna\footnote{\href{https://optuna.readthedocs.io/en/stable/}{Optuna} is a hyperparameter optimization framework that offers high flexibility and parallel computing.}. The search algorithm optimizes the following parameters:

\begin{enumerate}
	\item Sliding window size
	\item Feature type
	\item Target type and duration
	\item Measure of dependence
\end{enumerate}

Referring to Figure \ref{fig_data}, the sliding window step size $N$ is fixed at 100 days, while $W$ represents the sliding window size in the range of 50 to 200 days. For each point in time $t_i$, the measure of dependence is calculated. This enables the identification of regime changes in the data. For instance, it could happen that during a downtrend, different features might prevail than during an uptrend. Therefore, a trending indicator is calculated for each time step to show if the stock price will go $DOWN$, $UP$, or be $NEUTRAL$.

\begin{figure*}[h]
	\caption{Parallel coordinate plot with selective sliders on the vertical axis to filter hyperparameters.}
	\centering
	\includegraphics[width=1.0\textwidth]{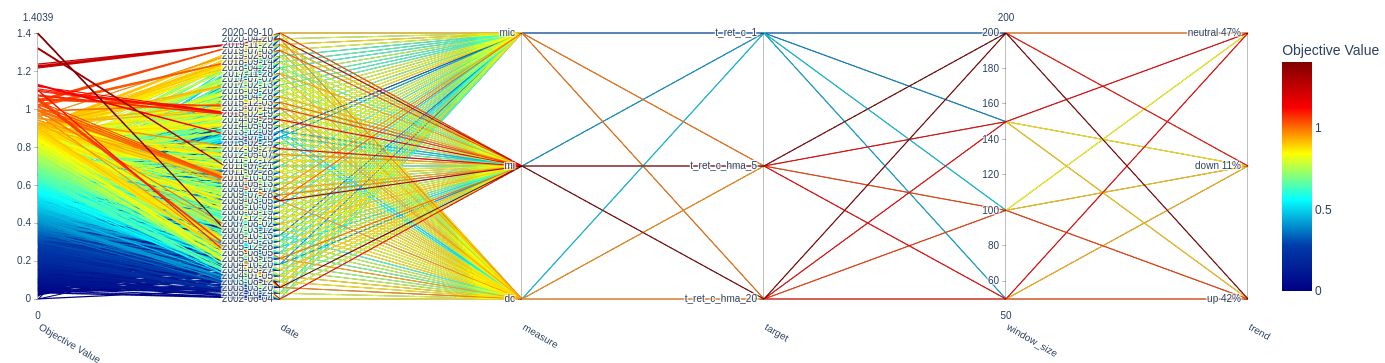}
	\label{fig_coordinate_plot}
\end{figure*}

Optuna offers various useful analysis tools, such as feature importance plots or parallel coordinate plots. With these tools, it is possible to find features with high predictive power, analyze the change of importance over time\footnote{For this purpose, the Optuna source code has been extended, integrating date functionality.} or verify which features prevail during an up or down trend. Figure \ref{fig_coordinate_plot} shows an example of parallel coordinate plots. Feature selection is done with the feature importance plots as it cannot be directly integrated into parallel coordinate plots.

The objective value is in this case the value of the measure of dependence. Based on the objective value the best sliding window size, features and targets can be selected. This parameters can then be used in the production model to verify if measure of dependence holds true, see \ref{chap_method}

The algorithm has been implemented in Python Sklearn \cite{scikit-learn} and Optuna. The Jupyter notebook can be found on \href{https://github.com/deeepwin}{Github} as a reference.

\subsection{Results}

In this chapter we give a summary of the results of the search process in table \ref{tab_results}. The results have not been verified using the production model, but shall serve as a first indication of what hyperparameters could be of importance.

\begin{table}[htbp]
	\caption{Hyperparameter search results for YUM stock.}
	\begin{center}
\begin{tabular}{|ll|l|l|l|}
	\hline
	\multicolumn{2}{|l|}{Hyperparameter} & MI & DC & MIC \\ \hline
	\multicolumn{1}{|l|}{\multirow{2}{*}{Window Size}} & Best & 50 & 50, 150, 200 & 50 \\ \cline{2-5} 
	\multicolumn{1}{|l|}{} & Worst & 50,150,200 & 150, 200 & 150,200 \\ \hline
	\multicolumn{1}{|l|}{\multirow{2}{*}{Features}} & Best & \begin{tabular}[c]{@{}l@{}}f\_ret\_c\_hma\_5\\ f\_ret\_c\_hma\_20\end{tabular} & f\_obv & \begin{tabular}[c]{@{}l@{}}f\_macd\\ f\_macd\_signal\end{tabular} \\ \cline{2-5} 
	\multicolumn{1}{|l|}{} & Worst & \begin{tabular}[c]{@{}l@{}}f\_ret\_c\_1\\ f\_vol\_pct\end{tabular} & \begin{tabular}[c]{@{}l@{}}f\_ret\_c\_1\\ f\_vol\_pct\end{tabular} & f\_ret\_c\_1 \\ \hline
	\multicolumn{1}{|l|}{\multirow{2}{*}{Targets}} & Best & \begin{tabular}[c]{@{}l@{}}t\_ret\_c\_hma\_5\\ t\_ret\_c\_hma\_20\end{tabular} & \begin{tabular}[c]{@{}l@{}}t\_ret\_c\_hma\_5\\ t\_ret\_c\_hma\_20\end{tabular} & \begin{tabular}[c]{@{}l@{}}t\_ret\_c\_hma\_5\\ t\_ret\_c\_hma\_20\end{tabular} \\ \cline{2-5} 
	\multicolumn{1}{|l|}{} & Worst & t\_ret\_c\_1 & t\_ret\_c\_1 & t\_ret\_c\_1 \\ \hline
\end{tabular}
		\label{tab_results}
	\end{center}
\end{table}

Interestingly, all three measures of dependence chose a window size of 50 as the best. This implies that, according to the production model, a larger historic window of more than 50 days might not be necessary. All three measures indicate that daily price return (f\_ret\_c\_1) is actually not a good feature, which is surprising. On the other hand, the best features strongly depend on the measure itself. MIC appears to strongly emphasize MACD, whereas DC found the on-balance volume oscillator to be the best. The targets show a consistent pattern among all three measures. Using daily price returns (t\_ret\_c\_1) as a target does not allow finding high predictive power. Instead, using Hull Moving Averages of 5 and 20 days yields the best results, which is an interesting finding.

In a follow-up experiment, one could test shorter window sizes to check if even shorter window sizes can be used. However, it must be understood that reducing the window size decreases the number of samples in a single calculation. Based on experience, the sample size should not be smaller than 100.

Additionally, the experiment can be conducted with other stocks. It would be interesting to see if the results correlate between different stocks. Based on the results of this experiment, it would be interesting to conduct a sensitivity analysis with daily price returns and compare them to Hull Moving Averages. Clearly, the production model would achieve higher accuracy using averages than daily returns. By following the verification process according to Figure \ref{fig_verification}, it is possible to iteratively preselect important hyperparameters with fairly simple measures of dependencies such as MI, DC, or MIC.

\section{Conclusions}

One of the key findings of this work is the crucial need for careful preparation of time-series data. It is essential that time-series data align with the chosen measure of dependence and the specific problem statement. For example, correlating the entire dataset of a feature with a target value identifies dependencies, but these dependencies may span any periods of time. However, this approach does not yield information about predictive power, rendering it ineffective for predictive applications. Conversely, correlating historical data with future data, without information leakage, enables a measure of dependence to genuinely uncover predictive power. Hence, in this study, we propose utilizing an extensive time horizon of data for comparison with historical data to identify patterns or regimes in the dataset. The assumption is that these patterns and regimes form the core of identifying predictive power.

Another crucial finding is that, irrespective of the chosen measure of dependence, the model based on the measure of dependence will always differ from the production model. Typically, a measure of dependence employs a simple module, like a statistical model. In contrast, a production model utilizes a more complex model, such as a deep learning network. When the measure of dependence employs a model as complex as a deep learning network, the feature engineering step for preselecting features becomes redundant. Instead, it is more advisable to directly evaluate hyperparameters with the production model, even though this incurs a high computational cost. Depending on the available infrastructure and cost budget, it is preferable to use a simple measure of dependence. Additional efforts to synchronize the measure and production model may be required. This work introduces a verification method to facilitate such synchronization. The method can be executed manually or automated for time-saving.

Finally, the experiments in this work have revealed some intriguing findings. Although they still require verification, daily price returns seem to be an unfavorable choice for the application of identifying predictive power in stock price information. Instead, using average data, especially employing Hull Moving Averages, which utilizes longer time periodicity, reveals higher predictive power.

\newpage
\bibliographystyle{ieeetr}
\bibliography{refs}

\end{document}